\documentclass[twoside]{article}

\usepackage{graphicx}
\usepackage{espcrc2mod}

\begin{document}
\sloppy

\begin{titlepage}
\title{Description of elastic vectormeson production and $F_2$ by two pomerons\thanks{Talk presented at {\it QCD98} in Montpellier, France. To be published in the proceedings {\it Nucl.Phys.B(Proc.Suppl.).}}}
\author{Michael Rueter\thanks{supported by a MINERVA-fellowship}
\address{School of Physics and Astronomy, Department of High Energy Physics, Tel-Aviv University, 69978 Tel-Aviv, Israel\\e-mail: {\tt rueter@post.tau.ac.il}}}
\begin{abstract}
  Using the Model of the Stochastic Vacuum many diffractive processes
  have been calculated by investigating the dipole-dipole scattering
  at a cm-energy of 20 GeV. In this work we extend the calculation to
  larger energies and small dipoles. We assume that there are two
  pomerons, the hard- and the soft-pomeron, which cause the different
  energy dependence for processes dominated by small or large dipoles.
  The physical processes are obtained by smearing the dipole-dipole
  amplitude with wavefunctions. For small dipoles the leading
  perturbative contribution is taken into account. By that way we can
  describe in addition to the already calculated low energy results
  (20 GeV) also the HERA data for the considered processes in nearly
  the whole energy and $Q^2$ range.
\end{abstract}
\end{titlepage}
\maketitle
\section{Method}
In this very short note we can only present some results for the $Q^2$ and energy dependence of elastic vectormeson production and the proton structure function $F_2$ and have to refer for technical steps, discussions and comparison with related works to our original publication \cite{Rueter:1998II}. The building block of our calculation is the dipole-dipole scattering amplitude. It is calculated in the framework of the Model of the Stochastic Vacuum (MSV) \cite{Dosch:1987,Dosch:1988}. Within this framework elastic hadron-hadron scattering \cite{Dosch:1994,Rueter:1996,Rueter:1996III}, hadron-dipole scattering \cite{Rueter:1997II}, photo- and electroproduction of vectormesons \cite{Dosch:1997,Kulzinger:1998} and $\pi^0$ \cite{Rueter:1998} and the proton structure function $F_2$ \cite{Dosch:1997II} were calculated but the cm-energy was always fixed at $20\; \mbox{GeV}$. We now extend this approach to higher energies by coupling a soft-pomeron, which trajectory is given by the Donnachie-Landshoff parameterization \cite{Donnachie:1992}, to large dipoles (both dipoles larger than a new introduced cut $c$) and a hard-pomeron with an intercept of 1.28 and a vanishing slope to small dipoles (at least one dipole smaller than $c$). Because the MSV can not be used for very small dipoles we cut these contributions if one of the dipoles is smaller than a second new cut $r_{\rm cut}$. By that way we can already describe the data for not to large $Q^2\le 35\;\mbox{GeV}^2$. In this regime one observes the transition from the soft to the hard behavior and it is the regime of our main interest. If we want to extend our approach to even harder processes we include for very small dipoles (smaller than $r_{\rm cut}$) the leading perturbative contribution with a running coupling which is frozen in the infra-red to be $\alpha_{\rm s}(\infty)=0.75$. By smearing this dipole-dipole scattering with process dependent wavefunctions, which we take from the cited references, we obtain different effective energy dependencies due to the different dipole sizes mainly involved. The two cuts $c$ and $r_{\rm cut}$ are the important new parameters. We did not perform a fit to the data but started with physical values and adjusted a little bit with the result: $c=0.35\;\mbox{fm},\;r_{\rm cut}=0.16\;\mbox{fm}$.
\section{Results}
\subsection{Vectormeson production}
\begin{figure}[ht]
\begin{minipage}{7.5cm}
\includegraphics[width=7.5cm]{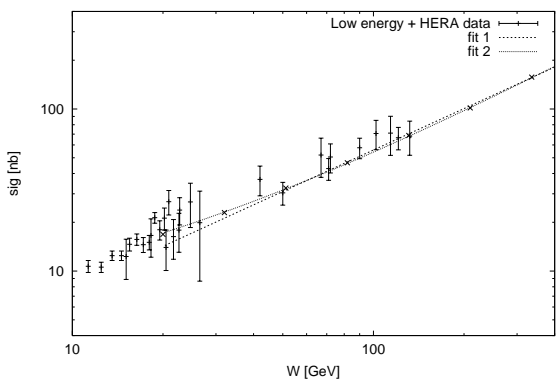}
\end{minipage}
{\vspace*{-1cm}
\caption{\scriptsize Our result (crosses) for the total elastic cross section of photoproduction of $J/\Psi$ as compared to data \protect\cite{Binkley:1982,Frabetti:1993,Derrick:1995II,Aid:1996II,Breitweg:1997II}. Fit 1 represents an exponential fit $\sigma^{\rm tot}=14.22\;\mbox{nb}\left(W/20\;\mbox{GeV}\right)^{0.85}$ and fit 2 is a fit with two powers which describes our results also for small $W$. There are also preliminary H1 data for larger $W$ (see for example reference \protect\cite{H1:1997}).}\label{photoJPsi}}
\end{figure}

\begin{figure}[ht]
\begin{minipage}{7.5cm}
\includegraphics[width=7.5cm]{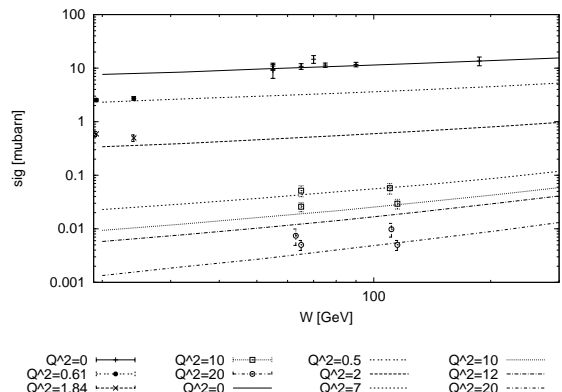}
\end{minipage}
{\vspace*{-1cm}
\caption{\scriptsize The total elastic cross section of $\rho$ production. The experimental data are from \protect\cite{Adams:1997,Aid:1996IIII,Derrick:1995IIII} for $Q^2=10{\;\rm GeV}^2$ and $20{\;\rm GeV}^2$. There are new preliminary HERA data (see for example \protect\cite{ZEUS:1997,H1:1997II}).}\label{electroRho}}
\end{figure}

All our old results, where only large dipoles contributed, are unchanged at $20\;\mbox{GeV}$ and we obtain an energy dependence through the soft-pomeron in agreement with experiment. An example is elastic hadron-hadron scattering or photoproduction of $\rho,\omega$ and $\phi$. For the $J/\Psi$ already small dipoles contribute and the energy dependence is much stronger (see Fig.\ref{photoJPsi}).
For electroproduction the effective pomeron slope increases with $Q^2$ as can be seen from Fig.\ref{electroRho} for the $\rho$. The simple exponential fit \mbox{$\sigma^{\rm tot}=a\left(W/20\;\mbox{GeV}\right)^b$} describes our result very good for large \mbox{$W>80\;\mbox{GeV}$} and we obtain
\[ \scriptsize
\begin{tabular}{|c|c|c|c|c|c|c|}
\hline
$Q^2[\mbox{GeV}^2]$&0.5&2&7&10&12&20\\
\hline\hline
$a[{\rm nb}]$&2064&291&17.5&7.25&4.43&1.08\\
\hline
$b$&0.34&0.44&0.71&0.77&0.82&0.92\\
\hline
\end{tabular}
\]
The rising of $b$ with $Q^2$ shows that by increasing the virtuality one probes smaller and smaller dipoles which are coupled to the hard-pomeron. This rise of the effective pomeron power with $Q^2$ is in agreement with the experiment. For the $J/\Psi$ this effect is smaller since already for photoproduction the dipoles are quite small. 
\subsection{$F_2(x,Q^2)$ and the total cross section of $\gamma$-$p$ scattering}
\begin{figure}[ht]
\leavevmode
\begin{minipage}{3.7cm}
\includegraphics[width=3.7cm]{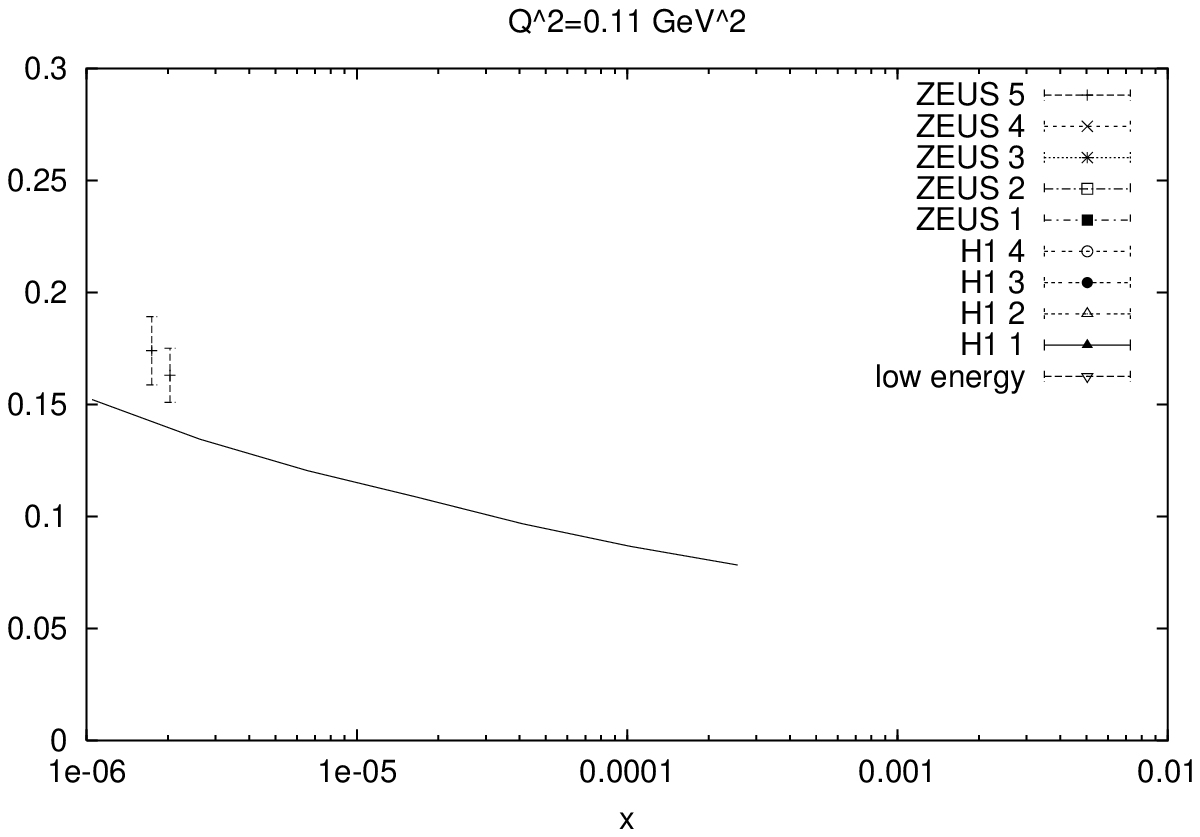}
\end{minipage}
\begin{minipage}{3.7cm}
\includegraphics[width=3.7cm]{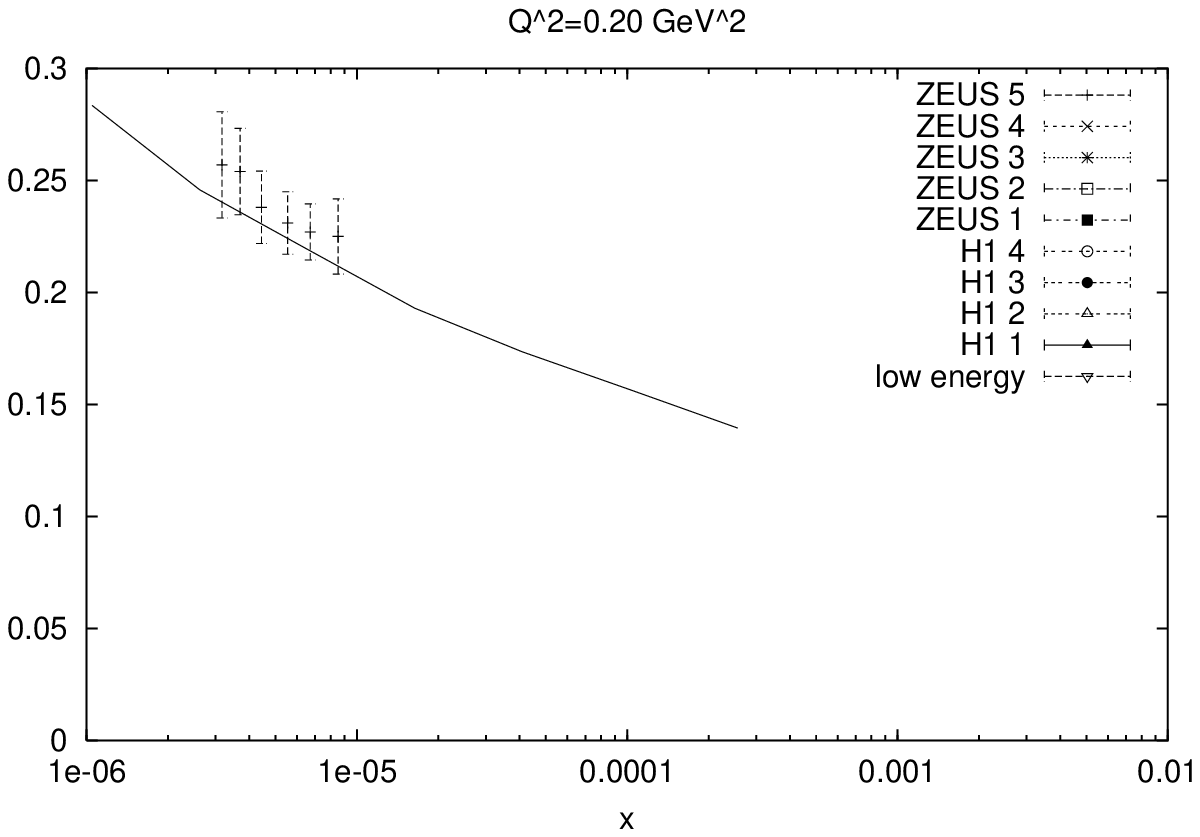}
\end{minipage}
\begin{minipage}{3.7cm}
\includegraphics[width=3.7cm]{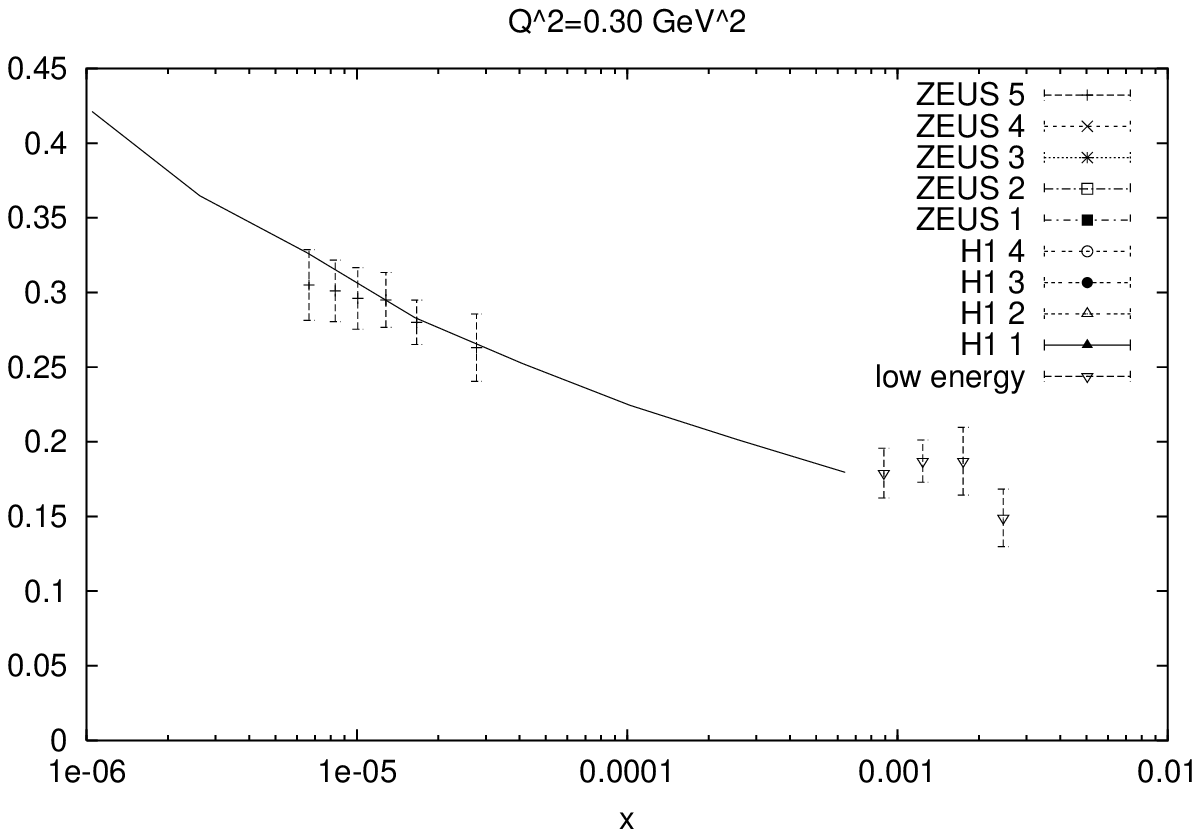}
\end{minipage}
\begin{minipage}{3.7cm}
\includegraphics[width=3.7cm]{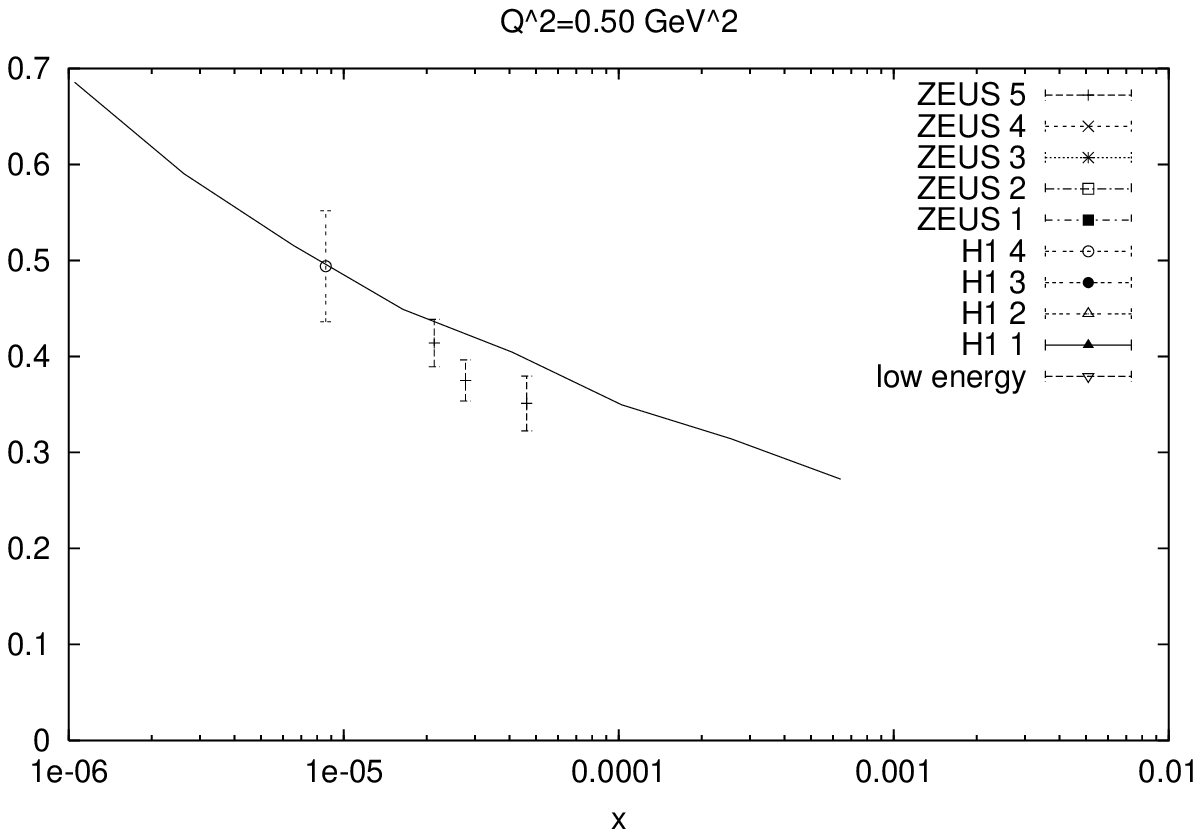}
\end{minipage}
\begin{minipage}{3.7cm}
\includegraphics[width=3.7cm]{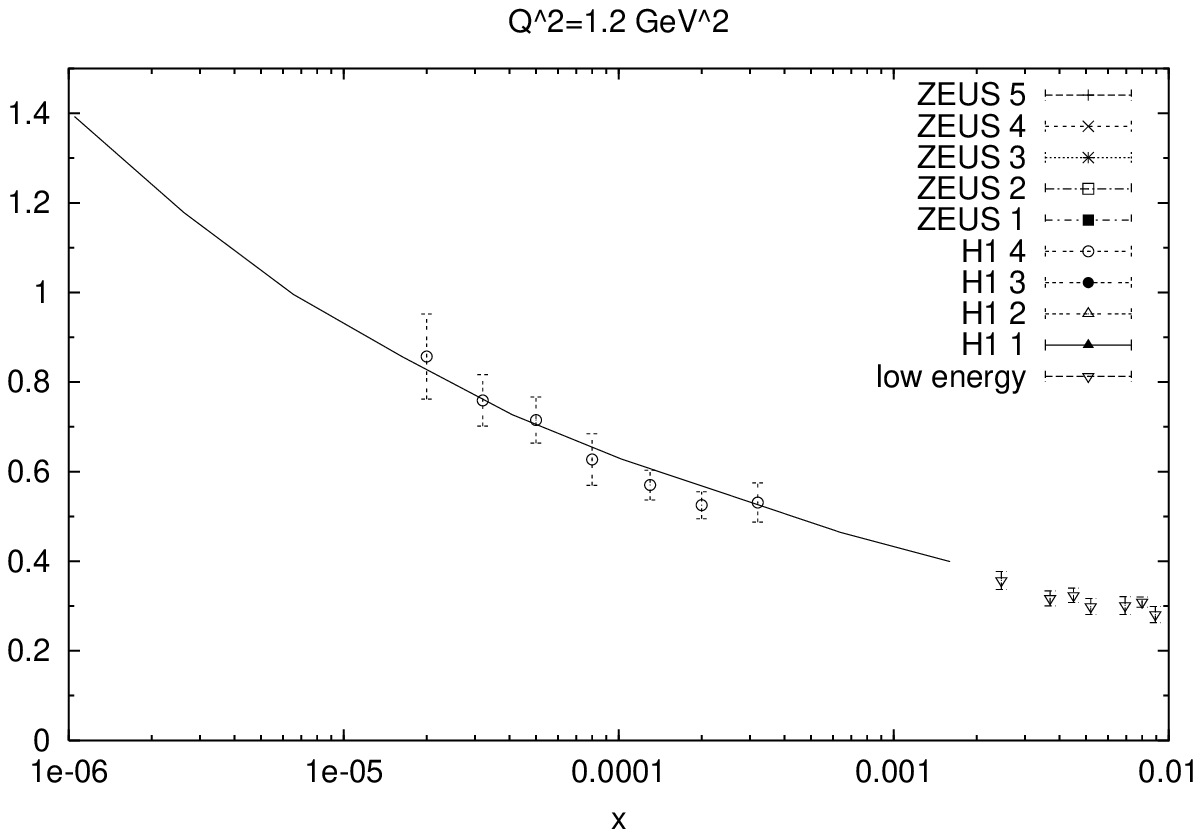}
\end{minipage}
\begin{minipage}{3.7cm}
\includegraphics[width=3.7cm]{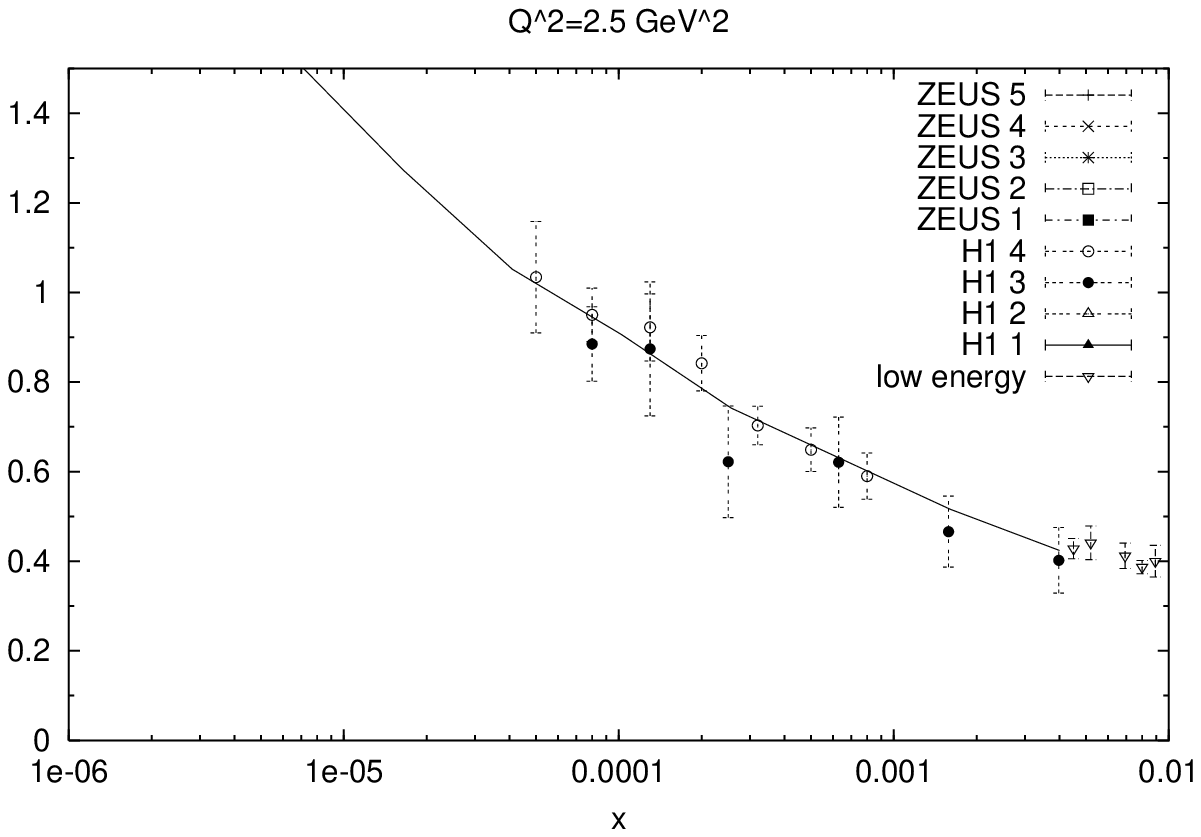}
\end{minipage}
\begin{minipage}{3.7cm}
\includegraphics[width=3.7cm]{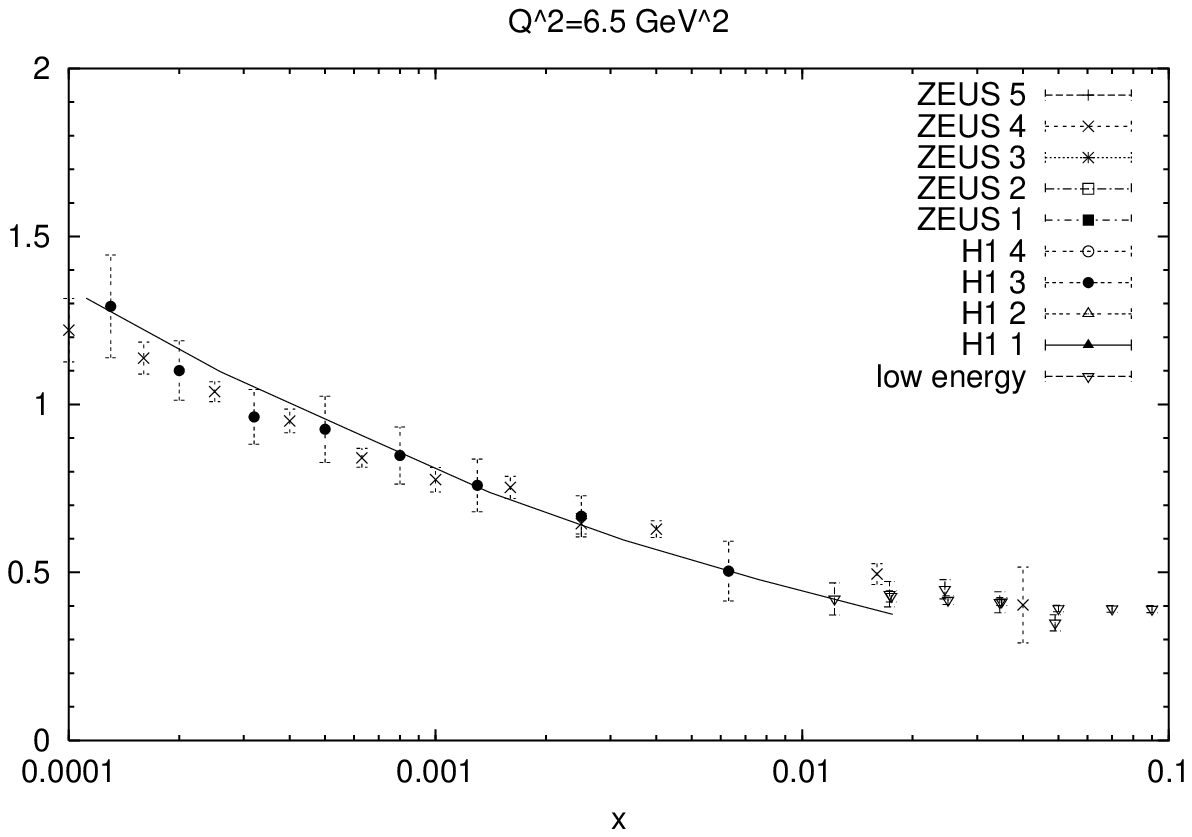}
\end{minipage}
\begin{minipage}{3.7cm}
\includegraphics[width=3.7cm]{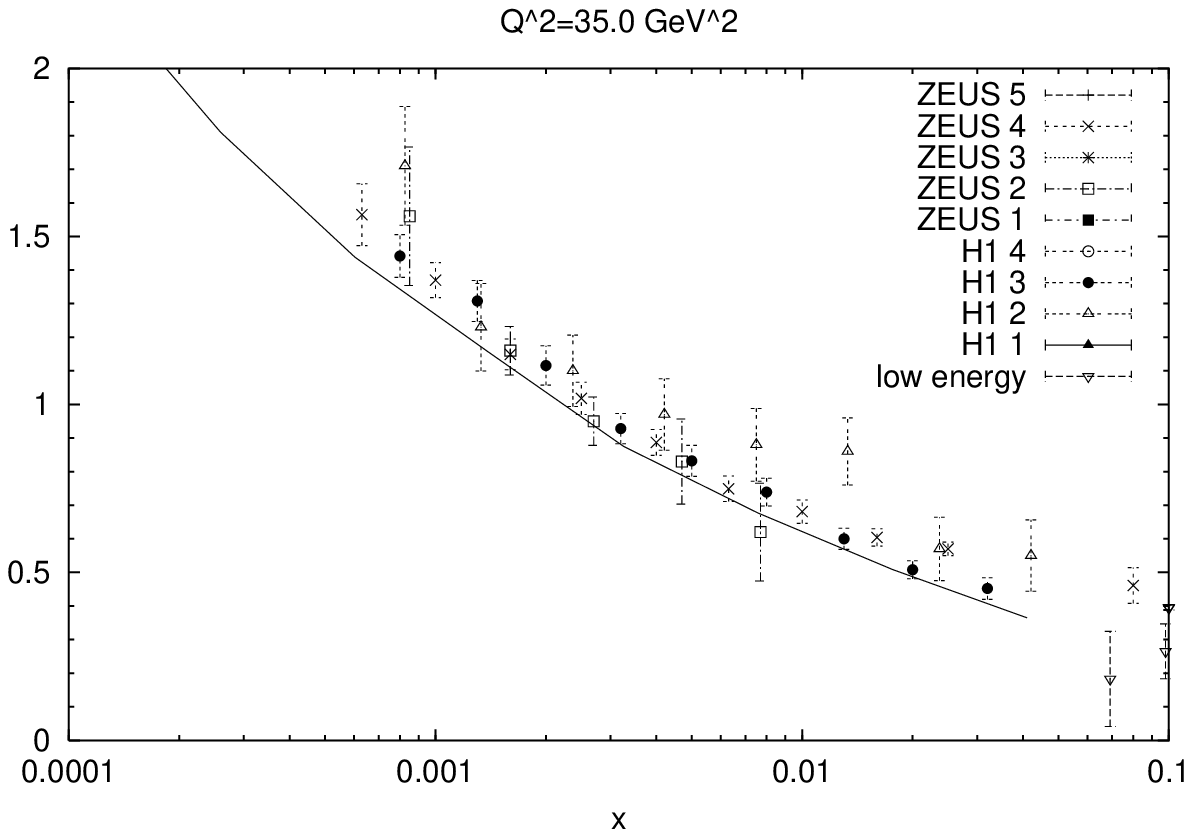}
\end{minipage}
\begin{minipage}{3.7cm}
\includegraphics[width=3.7cm]{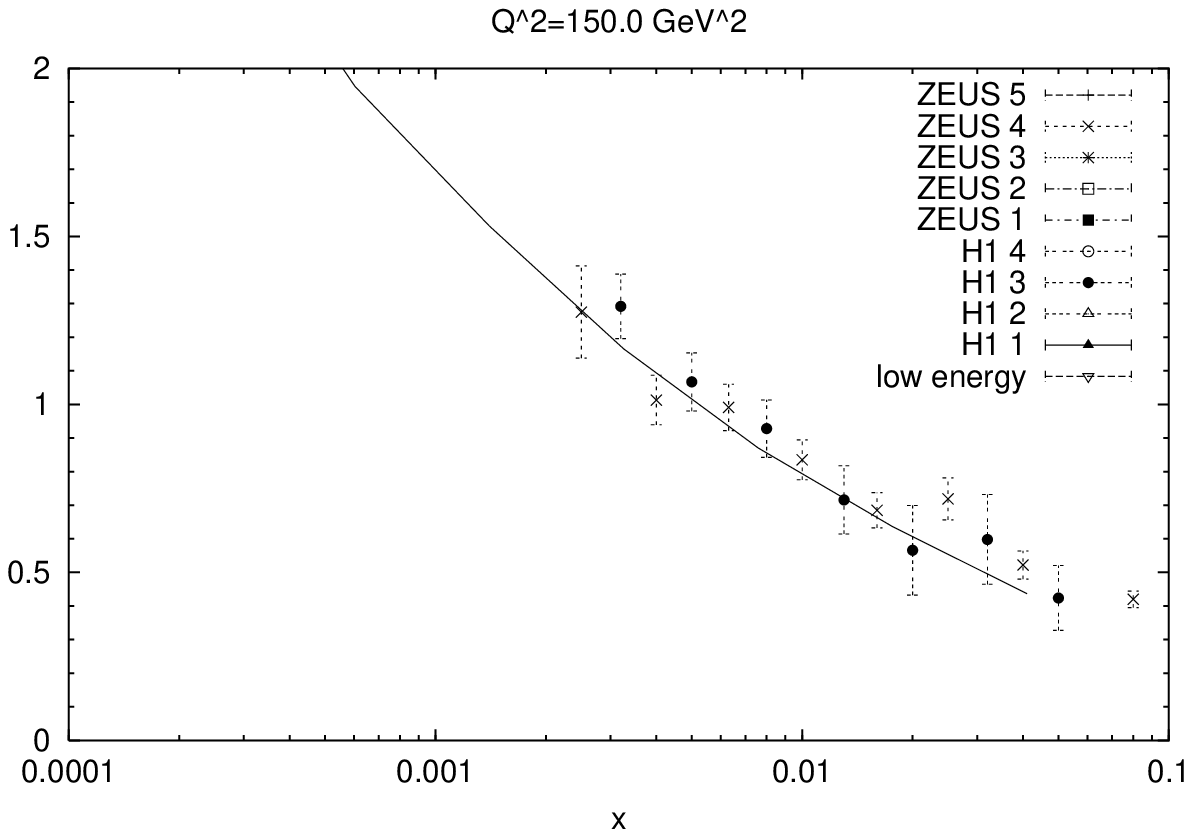}
\end{minipage}
\hfill
\begin{minipage}{3.7cm}
\includegraphics[width=3.7cm]{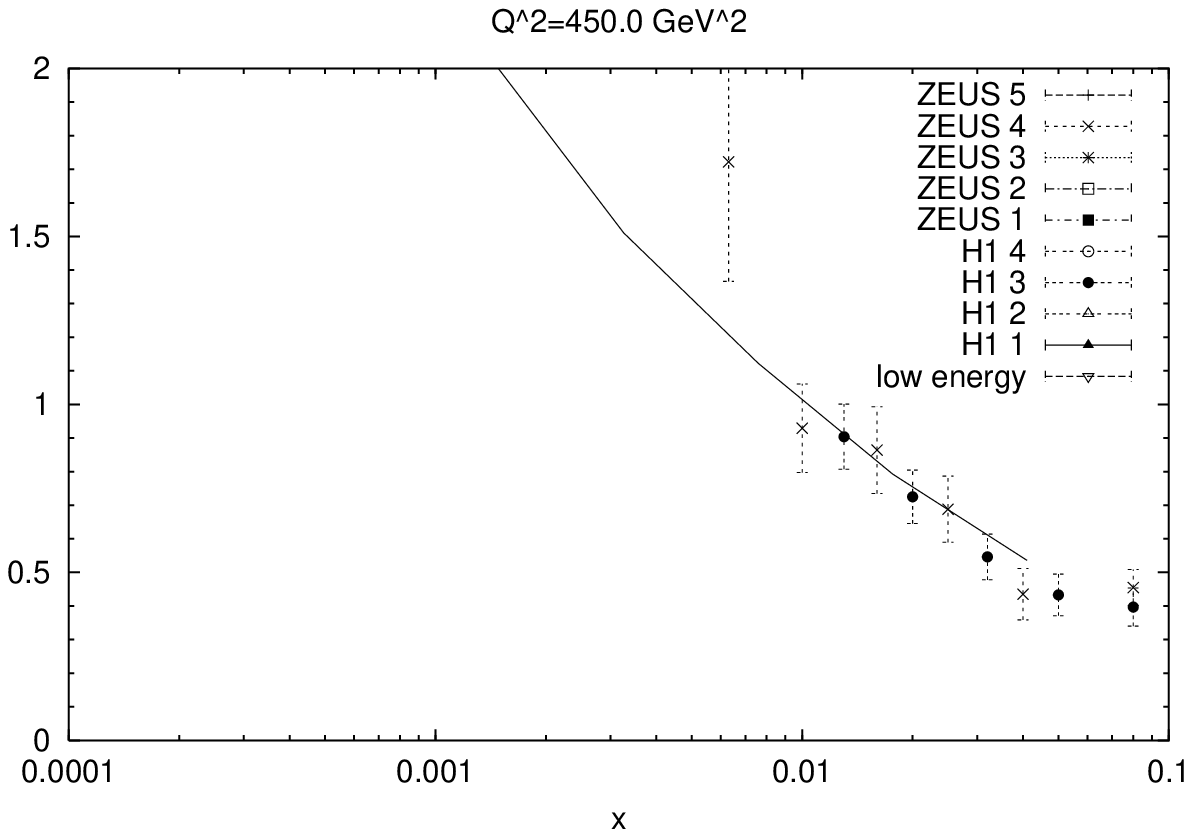}
\end{minipage}
\hfill
\begin{minipage}{3.7cm}
\includegraphics[width=3.7cm]{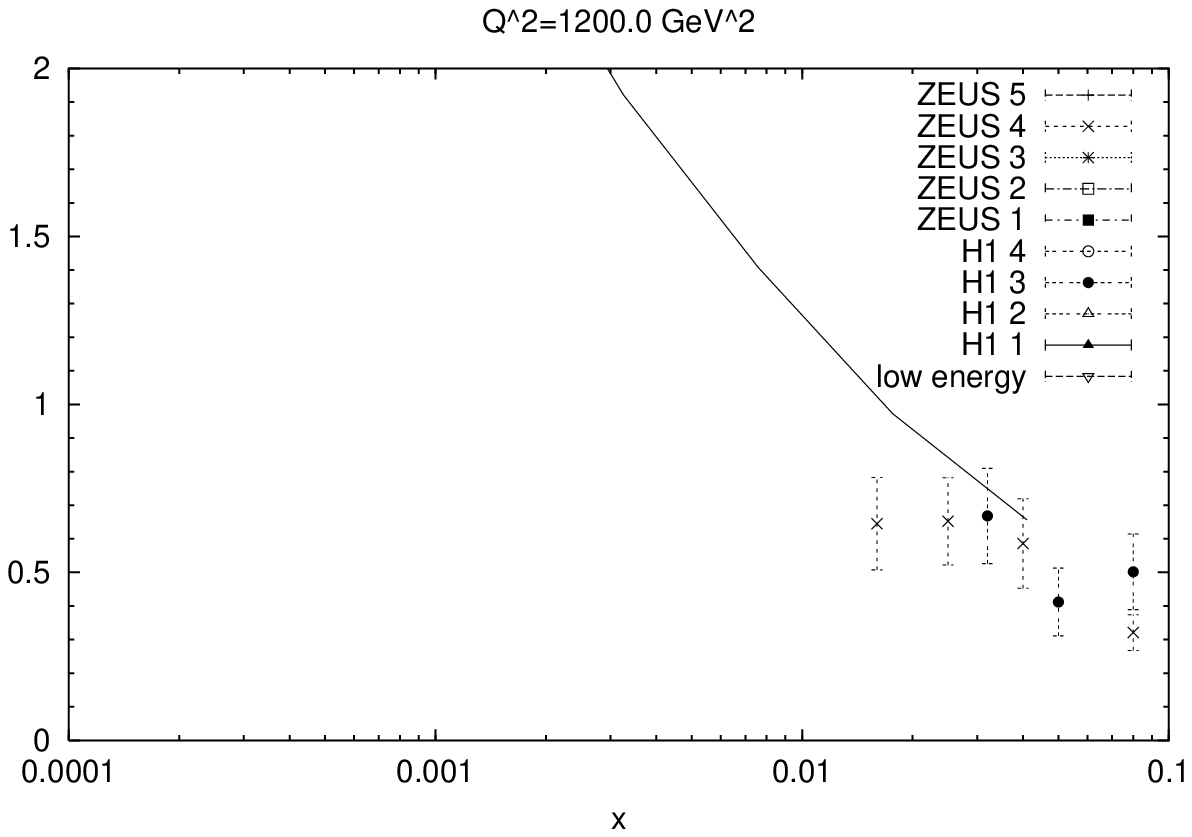}
\end{minipage}
\hfill
\begin{minipage}{3.7cm}
\includegraphics[width=3.7cm]{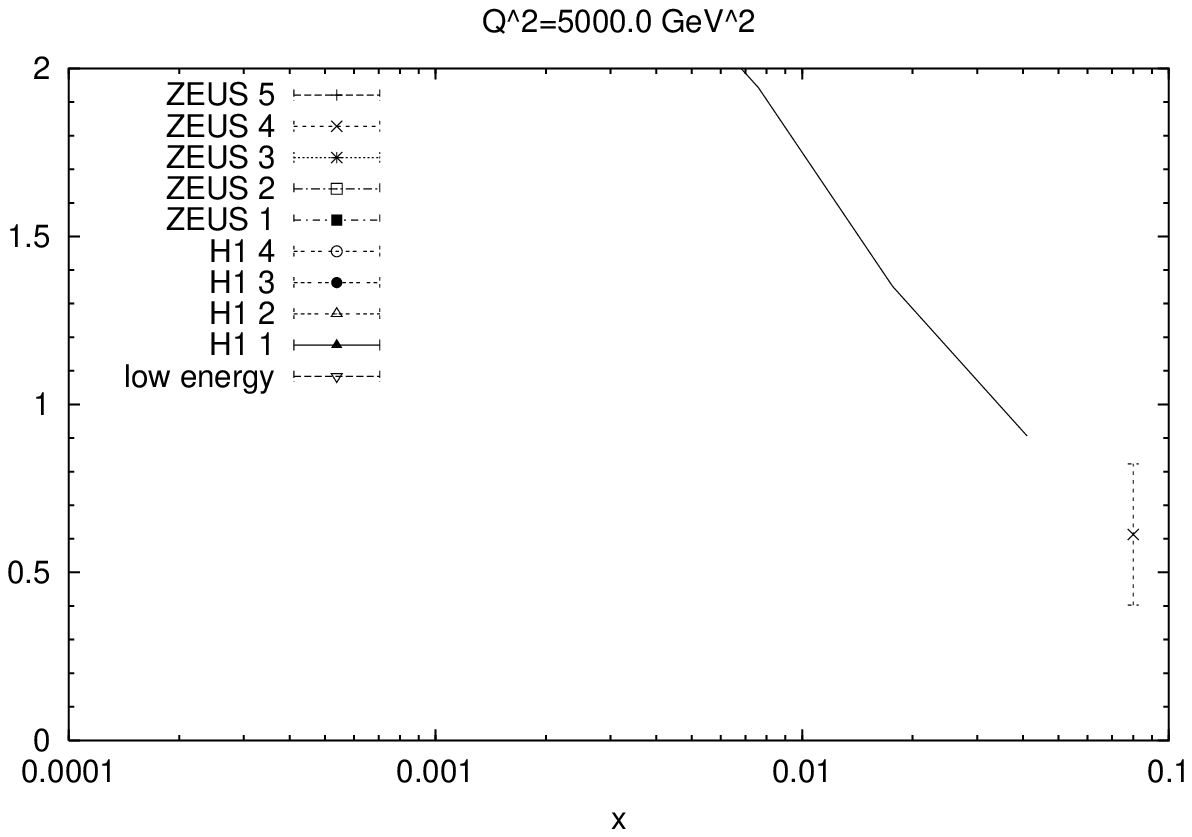}
\end{minipage}
{\vspace*{-1cm}
\caption{\scriptsize The proton structure function $F_2(x,Q^2)$ for fixed values of $Q^2$ as a function of $x$. The H1 data 1-4 are \protect\cite{Abt:1993,Ahmed:1995,Aid:1996III,Adloff:1997}, the ZEUS data 1-5 are \protect\cite{Derrick:1993,Derrick:1995III,Derrick:1996III,Derrick:1996IIII,Breitweg:1997III} and the low-energy data are \protect\cite{Benvenuti:1989,Arneodo:1997,Adams:1996}.}
\label{F2}}
\end{figure}

\begin{figure}
\leavevmode \begin{center}\includegraphics[width=7.5cm]{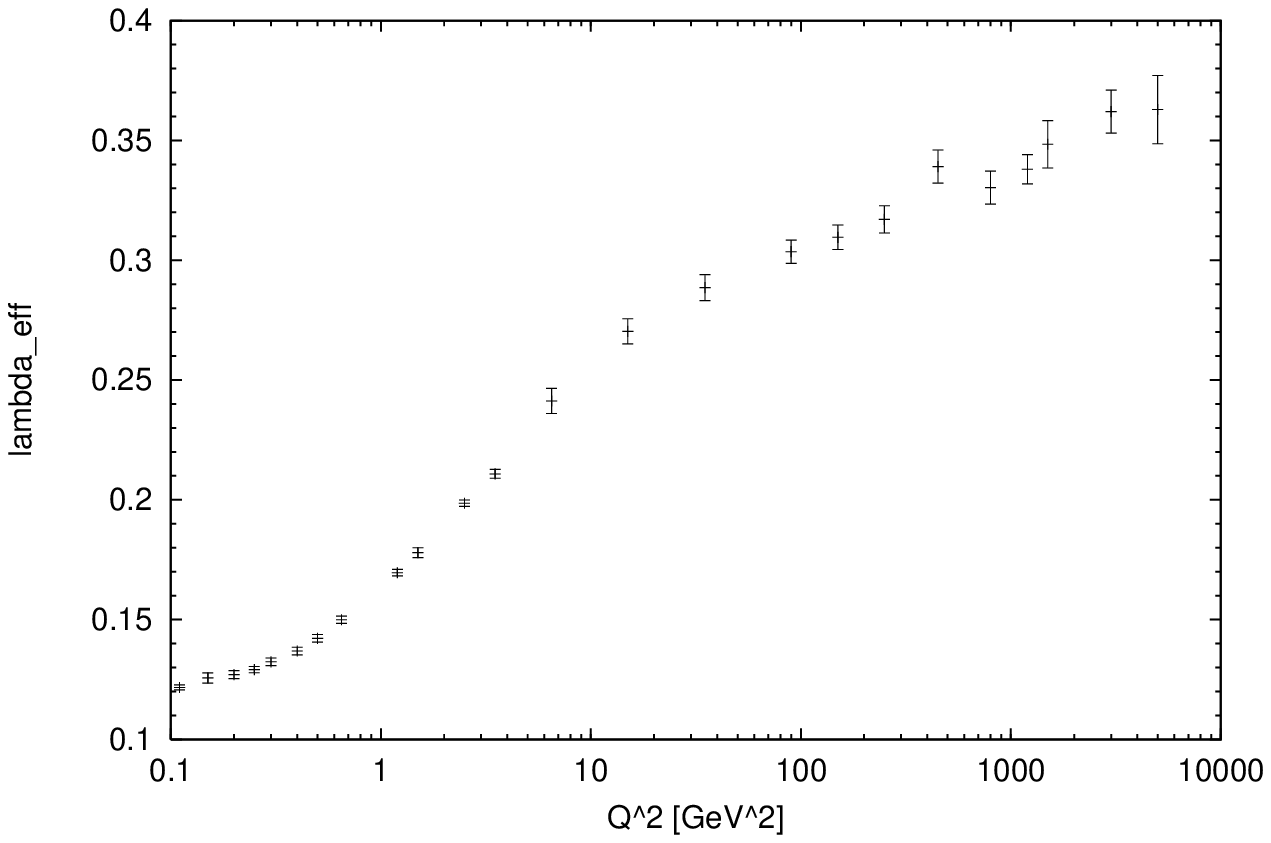}\hfill\includegraphics[width=7.5cm]{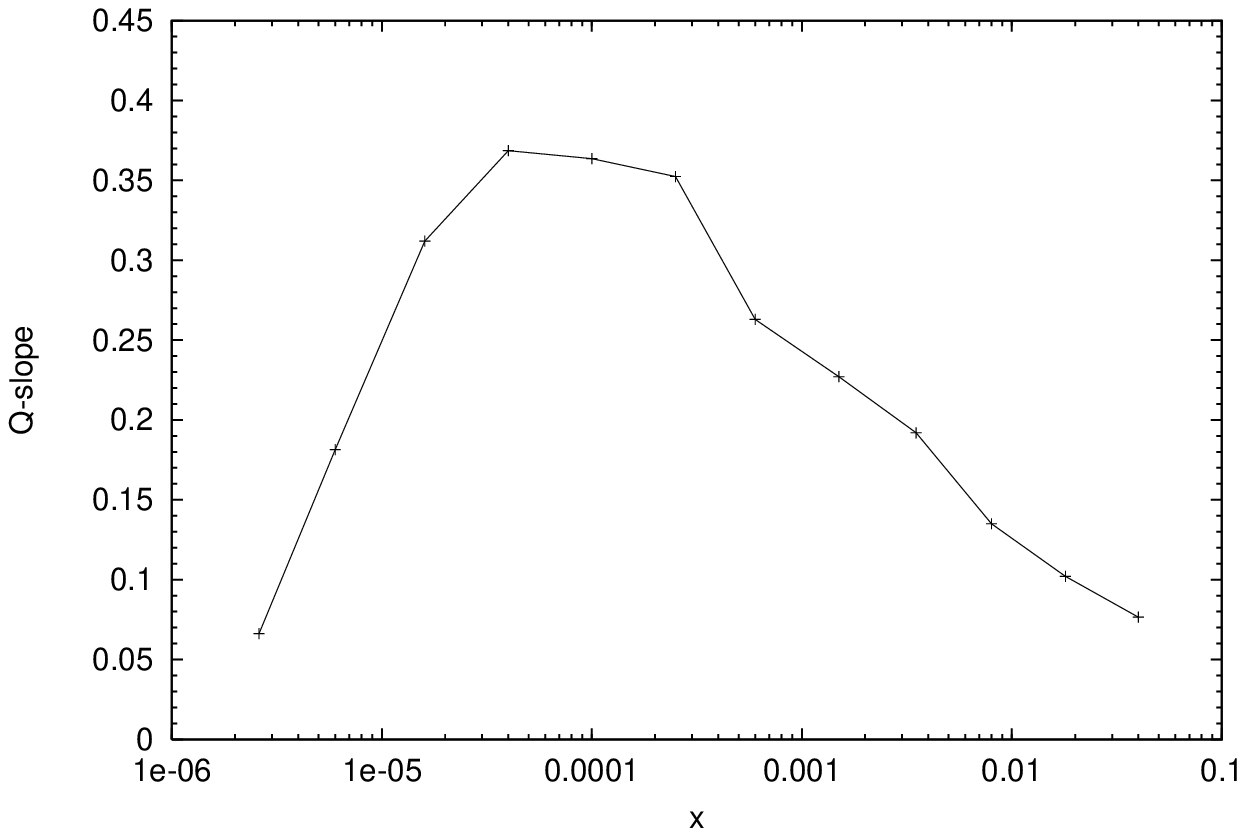}\end{center}
\unitlength1cm
{\vspace*{-1cm}
\caption{\scriptsize In the top the effective pomeron power $\lambda_{\rm eff}$ as a function of $Q^2$. The error bars are due to the numerical error of our results for $F_2$. In the bottom the Q-slope as a function of $x$.}
\label{lambdaeff}}
\end{figure}

\begin{figure}
\leavevmode \begin{center}\includegraphics[width=7.5cm]{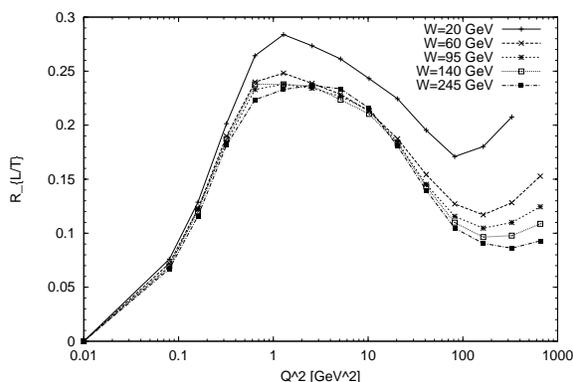} \end{center}
\unitlength1cm
{\vspace*{-1cm}
\caption{\scriptsize The ratio $R_{\rm L/T}$ for fixed $W$ as a function of $Q^2$. At $Q^2=0.01{\rm\;GeV}^2$ we put the photoproduction point.}
\label{Rlongtrans}}
\end{figure}

Using the different photon wavefunctions depending on the polarization we obtain the total cross sections $\sigma_{\rm L}$, $\sigma_{\rm T}$ and the proton structure function $F_2$ and $F_{\rm L}$ can be calculated
\begin{eqnarray}
F_2&=&\frac{1}{4\pi^2\alpha_{\rm em}}\frac{Q^4(1-x)}{Q^2+4m_{\rm P}^2 x^2}\left(\sigma_{\rm L}+\sigma_{\rm T}\right)\nonumber\\
F_{\rm L}&=&\frac{1}{4\pi^2\alpha_{\rm em}}\frac{Q^4(1-x)}{Q^2+4m_{\rm P}^2 x^2}\sigma_{\rm L}.\nonumber
\end{eqnarray}
The energy $W$ can be expressed by $W^2=Q^2/x-Q^2+m_{\rm P}^2$. In our approach we are limited to large energies, $W>20 \;\mbox{GeV}$, and small $x\le 0.05$ because we only take the pomeron and not the Regge contributions into account. In Fig.\ref{F2} we compare our result for $F_2$ with the experimental data for few values of $0.11 \;\mbox{GeV}^2\le Q^2 \le 5000\;\mbox{GeV}^2$.
The figures show that our model describes all the data in the limited $W$ and $x$ range very well. To obtain the effective energy dependence of the structure function for different scales we calculate the effective pomeron power $\lambda_{\rm eff}$ by fitting our results for $10^{-4}\le x\le 10^{-2}$ with
\[
F_2(x,Q^2)=a\frac{Q^4(1-x)}{Q^2+4m_{\rm P}^2 x^2}\left(\frac{Q^2}{x}-Q^2+m_{\rm P}^2\right)^{\lambda_{\rm eff}}
\]
for fixed $Q^2$. We also calculate the so called Q-slope
\[
\frac{\partial F_2(x,Q^2)}{\partial \ln Q^2}
\]
for fixed $x$ at values for $Q^2$ given like in reference \cite{Desgrolard:1998} by $Q^2_x=3.1\cdot 10^3\,x^{0.82}$ following the analysis of the experimental data in reference \cite{Caldwell:1997}. The result is shown together with the result for $\lambda_{\rm eff}$ in Fig.\ref{lambdaeff}.
Our results are in very good agreement with the experimental data \cite{Aid:1996III,Adloff:1997,Abramowicz:1997} and show the transition of the effective behavior going from large to small dipoles. Especially the Q-slope decreases for small $x$ as measured by the experiment. We also calculated the total $\gamma p$ cross section where we can include the photoproduction. We also investigated different contributions to $F_2$ like the charm contribution \cite{Rueter:1998II} or the longitudinal and transversal part by calculating the ratio of them as shown in Fig.\ref{Rlongtrans}.
\section{Acknowledgments}
I would like to thank H.G.~Dosch and Sandi Donnachie for many fruitful
discussions and suggestions. I am especially grateful to the {\it
  Minerva}-Stiftung for my fellowship and for supporting my conference visit.

\end{document}